\documentclass{article}
\usepackage[hyphens]{url}
\usepackage{dcase2019,amsmath,graphicx,url,times,booktabs, tabularx}
\usepackage[bookmarks=false, hidelinks]{hyperref}

\newcommand{\lwlrap}{\emph{l$\omega$lrap}}

\title{Audio tagging with noisy labels and minimal supervision}

\name{Eduardo Fonseca$^{1}\sthanks{This work is partially supported by the European Union's Horizon 2020 research and innovation programme under grant agreement No 688382 AudioCommons and a Google Faculty Research Award 2018.}\sthanks{Equal contribution.}$,
      Manoj Plakal$^{2}\footnotemark[2]$,
      Frederic Font$^{1}$, 
      Daniel P. W. Ellis$^{2}$,
      Xavier Serra$^{1}$
      }
\address{$^1$Music Technology Group, Universitat Pompeu Fabra, Barcelona \{name.surname\}@upf.edu\\          
        $^2$ Google, Inc., New York, NY, USA \{plakal,dpwe\}@google.com \\
 }
 
\begin{document}

\ninept
\maketitle

\begin{sloppy}

\begin{abstract}
This paper introduces Task 2 of the DCASE2019 Challenge, titled ``Audio tagging with noisy labels and minimal supervision''. This task was hosted on the Kaggle platform as ``Freesound Audio Tagging 2019''.
The task evaluates systems for multi-label audio tagging using a large set of noisy-labeled data, and a much smaller set of manually-labeled data, under a large vocabulary setting of 80 everyday sound classes.
In addition, the proposed dataset poses an acoustic mismatch problem between the noisy train set and the test set due to the fact that they come from different web audio sources.
This can correspond to a realistic scenario given the difficulty of gathering large amounts of manually labeled data.
We present the task setup, the FSDKaggle2019 dataset prepared for this scientific evaluation, and a baseline system consisting of a convolutional neural network. All these resources are freely available.
\end{abstract}


\section{Introduction}
\label{sec:intro}
Environmental sound recognition has gained attention in recent years, encompassing tasks such as acoustic scene classification, sound event detection or audio tagging \cite{virtanen2018computational}.
The latter is becoming a popular task partly due to the various audio tagging tasks in the DCASE Challenge editions, and its impact on applications such as automatic description of multimedia, or acoustic monitoring.
This paper describes the characteristics, dataset and baseline system of DCASE2019 Task 2 ``Audio tagging with noisy labels and minimal supervision''.

Everyday sound tagging consists of identifying the sound events present in an audio recording.
The most common approach to create sound event taggers relies on supervised learning through labeled audio datasets.
New released datasets tend to be of increasing size in order to allow exploitation of data-driven approaches, e.g., deep learning.
However, manual labeling of large datasets is expensive and typically a limiting factor in machine listening; hence, creators are often forced to compromise between dataset size and label quality.
Thus, most recent datasets feature larger sizes \cite{gemmeke2017audio,fonseca2018general,Fonseca2019learning}, but their labeling is less precise than that of conventional small and exhaustively labeled datasets \cite{salamon2014dataset,piczak2015esc,foster2015chime}.
We are, therefore, witnessing a transition towards larger datasets that inevitably include some degree of label noise.
Likewise, the current trend is moving towards general-purpose sound event recognizers, able to recognize a broad range of everyday sounds.
This is favoured by the appearance of the AudioSet Ontology; a hierarchical tree with 627 classes encompassing the most common everyday sounds \cite{gemmeke2017audio}.
This implies that not only are we interested in recognizing typical sound sources (e.g., \emph{Bark}), but also less usual sound classes (e.g., production mechanisms such as \emph{Fill (with liquid)}).
Further, not only are these classes less frequent, but also some can be semantically or acoustically similar to others (e.g. \emph{Trickle, dribble} and \emph{Fill (with liquid)}).
Manual annotation of these more ambiguous categories becomes more difficult,
which makes them more prone to labeling errors.
An alternative to gathering data for training general-purpose audio taggers is to retrieve audio and metadata from websites such as Freesound or Flickr.
Labels can be inferred automatically by using automated heuristics applied to the metadata, or applying pre-trained classifiers on the audio material.
This approach supports rapid collection of large amounts of data, but at the cost of a considerable amount of label noise.

In this context, label noise arises as a challenge in general-purpose sound event recognition, including adverse effects such as performance drop or increased complexity of models \cite{frenay2014classification}, and also hindering proper learning of deep networks~\cite{arpit2017closer,zhang2016understanding}.
Consequently, coping with label noise could open the door to better sound event classifiers, and could allow the exploitation of large amounts of web audio for training, while reducing manual annotation needs.
The topic of \textit{learning with noisy labels} is a consolidated research area in computer vision \cite{han2018co,malach2017decoupling,jiang2017mentornet,goldberger2016training,veit2017learning,zhang2018generalized,patrini2017making}.
However, in sound recognition it has received little attention, probably due to the conventional paradigm of learning from small and clean datasets; only a few works directly address the analysis and mitigation of label noise \cite{Fonseca2019learning, kumar2018learning, Fonseca2019model}. 

In this paper, we propose a task, a dataset and a baseline system to foster label noise research in general-purpose sound event tagging.
We follow up on DCASE2018 Task 2 \cite{fonseca2018general}, and propose to investigate the scenario where a small set of manually-labeled data is available, along with a larger set of noisy-labeled data, in a multi-label audio tagging setting, and using a vocabulary of 80 classes of everyday sounds.
The proposed task addresses two main research problems.
The first problem is how to adequately exploit a large quantity of noisy labels, many of which are incorrect and/or incomplete, and how to complement it with the supervision provided by a much smaller amount of reliable manually-labeled data (\textit{minimal} supervision).
The second problem is given by the acoustic mismatch between the noisy train set and the test set.
Distribution shifts between data have been shown to cause substantial performance drops in machine learning, both for vision \cite{recht2019imagenet} and audio \cite{mesaros2018acoustic}.
In our case, the noisy train set comes from a different web audio source than the test set, which is sometimes a real-world constraint.
This paper is organized as follows. 
Section \ref{sec:setup} provides more details about the task and its experimental setup.
Section \ref{sec:dataset} presents the dataset prepared for the task, and Section \ref{sec:baseline} describes a baseline system.
Final remarks are given in Section \ref{sec:conclusion}.

\section{Task Setup}
\label{sec:setup}
The goal of this task is to predict appropriate labels for each audio clip in a test set. 
The predictions are to be done at the clip level, i.e., no start/end timestamps for the sound events are required.
Some test clips bear one ground truth label while others bear several labels.
Hence, the task setup is a \emph{multi-label} classification problem and the systems to be developed can be denoted as multi-label audio tagging systems, as illustrated in Fig.~\ref{fig:task_diagram}. 
This task was hosted on the Kaggle platform from April 4th to June 10th 2019.
The resources associated to this task (dataset download, submission, and leaderboard) can be found on the Kaggle competition page.\footnote{\url{https://www.kaggle.com/c/freesound-audio-tagging-2019}\\Note that the competition name on Kaggle is abbreviated from the full DCASE2019 Challenge task name to ``Freesound Audio Tagging 2019''.\label{footnote_Kaggle_page}}

\begin{figure}[ht]
  \centering
  \centerline{\includegraphics[width=0.82\columnwidth]{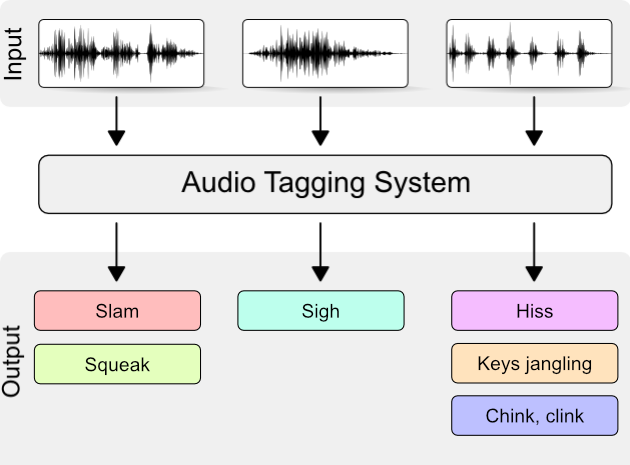}}
  \caption{Overview of a multi-label tagging system.}
  \label{fig:task_diagram}
\end{figure}
As described in Section~\ref{sec:dataset}, the audio data for this task consists of a test set and two train sets: a \emph{curated} train set and a \emph{noisy} train set, that allow to experiment with training data of different levels of reliability and coming from different sources.
System evaluation was carried out on Kaggle servers (see Section \ref{ssec:eval_metric}) using the test set, which is further split into two divisions, for the \textit{public} and \textit{private} leaderboards.
During the competition, the test subset corresponding to the public leaderboard was used to provide live ranking of all participants.
To compute the final private leaderboard, at the end of the competition, systems were re-evaluated using the unseen private test set, of which neither the audio nor the labels were accessible to participants. 

\subsection{Evaluation Metric and Competition Rules}
\label{ssec:eval_metric}
The task was evaluated via label-weighted label-ranking average precision (abbreviated as {\lwlrap} and pronounced ``lol wrap''). This generalizes the mean reciprocal rank (MRR) used in the 2018 challenge~\cite{fonseca2018general} to the case of multiple true labels per test item. 
Let $Lab(s, r)$ be the class label at rank $r$ (starting from 1) in test sample $s$, and $Rank(s, c)$ be the rank of class label $c$ in that list, i.e. $Lab(s, Rank(s, c)) = c$.
Then, if the set of ground-truth classes for sample $s$ is $C(s)$, the label-ranking precision for the list of labels up to class $c$ (assumed to be in $C(s)$) is:
\begin{equation}
    Prec(s, c) = \frac{1}{Rank(s, c)} \sum_{r=1}^{Rank(s, c)} \mathbf{1}[Lab(s, r) \in C(s)]
\end{equation}
where $\mathbf{1}[\cdot]$ evaluates to 1 if the argument is true, else zero. 
$Prec(s, c)$ is equal to 1 if all the top-ranked labels down to $c$ are part of $C(s)$, and at worst case equals $1/Rank(s, c)$ if none of the higher-ranked labels are correct.
In contrast to plain \emph{lrap}, which averages precisions within a sample then across samples, thereby downweighting labels that occur on samples with many labels, \emph{l$\omega$lrap} calculates the precision for each \emph{label} in the test set, and gives them all equal contribution to the final metric:
\begin{equation}
    l\omega lrap = \frac{1}{\sum_s |C(s)|} \sum_s \sum_{c \in C(s)} Prec(s, c)
\end{equation}
where $|C(s)|$ is the number of true class labels for sample $s$.
We use label weighting because it allows per-class values to be calculated, while keeping the overall metric as a simple average of the per-class metrics (weighted by each label's prior in the test set). A Python implementation of {\lwlrap} is provided in \footnote{\url{https://colab.research.google.com/drive/1AgPdhSp7ttY18O3fEoHOQKlt_3HJDLi8}}.

This scientific evaluation was set up as a Kaggle \emph{Kernels-only} competition.
This means that all participants had to submit their systems as inference models in Kaggle Kernels (similar to \emph{Jupyter Notebooks}), to be evaluated on remote servers.
In addition, inference run-time was limited to a maximum of one hour in a Kernel with one GPU, and memory constraints were also imposed.
These constraints aim to discourage the usage of large model ensembles.
Participants could submit a maximum of two submissions per day, and select two \emph{final} submissions to be considered for the private leaderboard ranking. 
A detailed description of the task rules can be found in the Rules section of the competition page;\textsuperscript{\ref{footnote_Kaggle_page}} the most important points are summarized in the DCASE Challenge page.\footnote{\url{http://dcase.community/challenge2019/task-audio-tagging\#task-rules}} 

To complement the leaderboard results of the {\lwlrap} ranking, the task organizers introduced a complementary Judges' Award to promote submissions using novel, problem-specific and efficient approaches. 
Details about the Judges' Award rules can be found in the Judges' Award section of \textsuperscript{\ref{footnote_Kaggle_page}}.

\vspace{-4mm}
\section{Dataset}
\label{sec:dataset}

The dataset used is called \emph{FSDKaggle2019}, and it employs audio clips from the following sources:
\begin{itemize}
  \item Freesound Dataset (FSD): a dataset under development based on Freesound content organized with the AudioSet Ontology \cite{gemmeke2017audio}. Freesound is a sound sharing site hosting over 400,000 clips uploaded by a community of users, who additionally provide some basic metadata, e.g., tags. \cite{font2013freesound}. These data are used to create the curated train set and the test set.
  \item The soundtracks of a pool of Flickr videos taken from the Yahoo Flickr Creative Commons 100M (YFCC100M) dataset \cite{YFCC100M}. These data are used to create the noisy train set.
\end{itemize}

FSDKaggle2019 is freely available from Zenodo\footnote{\url{https://doi.org/10.5281/zenodo.3612637}}, all clips are provided as uncompressed PCM 16 bit 44.1 kHz mono audio files, its ground truth labels are provided at the clip-level (i.e., weak labels), and its partitioning is depicted in Fig.~\ref{fig:data_split}.
\begin{figure}[ht]
   \vspace{-2mm}
    \centering
  \centerline{\includegraphics[width=\columnwidth]{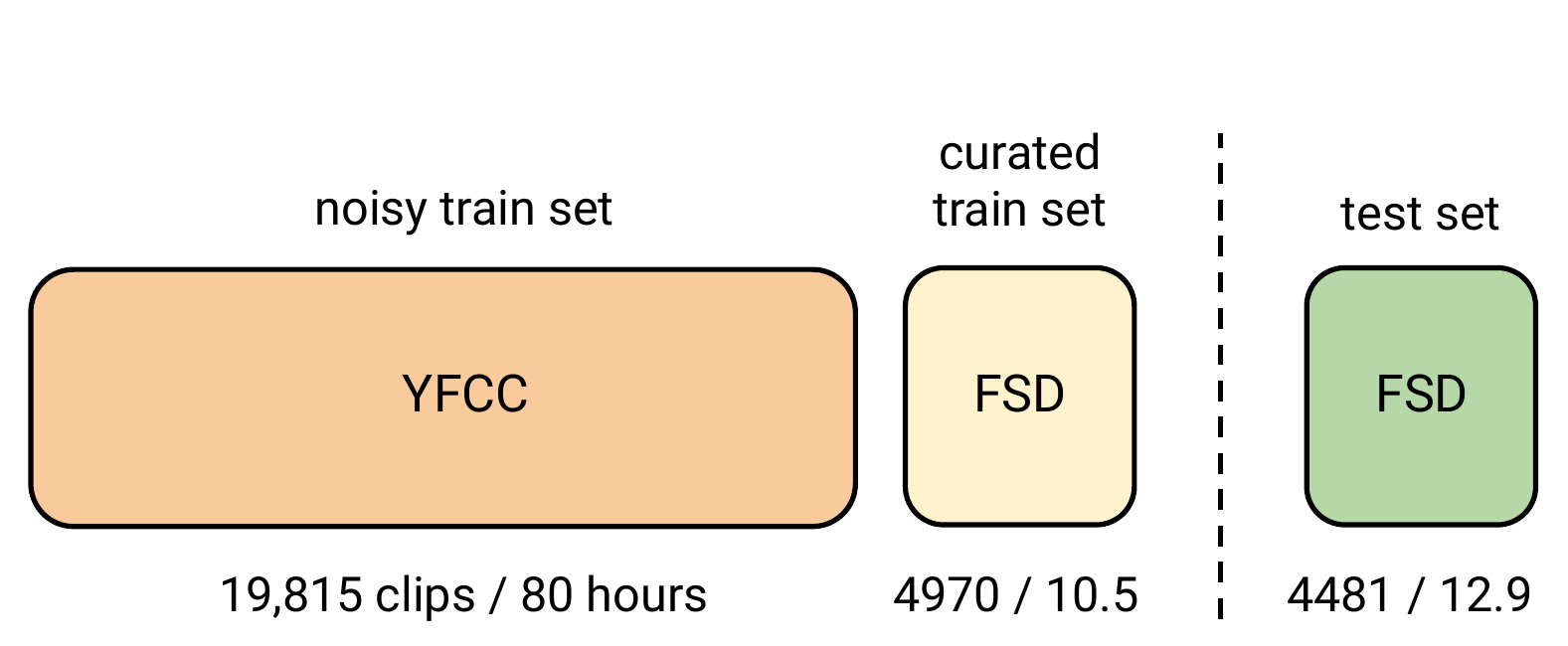}}
     \vspace{-1mm}
    \caption{Data split in FSDKaggle2019, including number of clips / duration in hours, and data origin. Colors depict quality of labels: orange, yellow and green correspond to noisy labels, correct but potentially incomplete labels, and exhaustive labels, respectively.}
    \label{fig:data_split}
\vspace{-2mm}
\end{figure}

\vspace{-1mm}
\subsection{Curated train set and test set}
\label{ssec:curated_train}
The first step carried out in the creation of FSDKaggle2019 was the definition of a vocabulary of 80 classes drawn from the AudioSet Ontology \cite{gemmeke2017audio}.
This vocabulary was chosen based on the following criteria: \textit{i)} we consider leaf nodes of the AudioSet hierarchy for which there is enough data available in FSD, \textit{ii)} we aim to encompass a diverse range of everyday sounds, and \textit{iii)} we remove few clearly isolated classes (those with the weakest semantic relations with any of the rest), thus promoting confounds between semantically/acoustically similar classes to some extent.
The main sound \textit{families} (i.e., groups of sound classes) in the resulting vocabulary are, in descending order of prevalence, human sounds, domestic sounds, musical instruments, vehicles, animal sounds, natural sounds, materials, and mechanisms.
The full list of 80 classes is available in the Data section of \textsuperscript{\ref{footnote_Kaggle_page}}. 


In a second step, we did a mapping of Freesound clips to the selected 80 class labels.
To this end, a set of keywords was defined connecting the user-provided Freesound tags with the AudioSet labels.
Using this mapping, for every class, we retrieved the audio clips that feature at least one of the defined keywords among their tags.
This process led to a number of automatically-generated \textit{candidate annotations} indicating the potential presence of a sound class in an audio clip (i.e., weak labels, as timing information is not included).
Nonetheless, in some audio clips the target signal fills the clip length almost completely, which can be considered as a strong label.
Subsequently, the candidate annotations were human-validated using a \textit{validation} task deployed in \emph{Freesound Annotator},\footnote{\url{https://annotator.freesound.org}\label{footnote_url_FSDs}} an online platform for the collaborative creation of open audio datasets \cite{Fonseca2017freesound}.
In this task, users verify the presence/absence of a candidate sound class in an audio clip with a \emph{rating} mechanism.
The vast majority of provided labels have inter-annotator agreement but not all of them.
The outcome is a set of clips where the corresponding label(s) are correct; nevertheless, it can happen that a few of these audio clips present additional acoustic material beyond the provided label(s).

The resulting data were split into a train set and a test set. 
We refer to this train set as \textit{curated} in order to distinguish it from the noisy set described in Section \ref{ssec:noisy_train}.
To mitigate train-test contamination, the split was carried out considering the clip uploaders in Freesound.
We allocated all audio clips uploaded from the same user into either the curated train set or the test set, so that the sets are disjoint at the Freesound user level.
The partition proportion was defined to limit the supervision provided in the curated train set, thus promoting approaches to deal with label noise.

Finally, the test set was further annotated using a label \textit{generation} tool \cite{favory2018facilitating}, in which \textit{i)} pre-existent labels can be re-validated, and \textit{ii)} potentially missing labels can be added through exploration of the AudioSet Ontology.
The outcome is a set of exhaustively labeled clips where the label(s) are correct and complete considering the target vocabulary; nonetheless, few clips could still present additional (unlabeled) acoustic content out of the vocabulary.

The main characteristics of the curated train set, noisy train set and test set are listed in Table~\ref{tab:dataset_stats}.
The curated train set consists of 4970 clips with a total of 5752 labels. 
Labels per clip ranges from 1 to 6 with a mean of 1.2.
The test set consists of 4481 clips with a total of 6250 labels. 
Labels per clip ranges from 1 to 6 with a mean of 1.4.
Note the increased number of labels per clip with respect to the curated train set, due to the process of exhaustive labelling.
In both cases, clip length ranges from 0.3s to 30 due to the diversity of the sound classes and the preferences of Freesound users when recording/uploading sounds.

\begin{table}[t!]
\vspace{-1mm}
\caption{Main stats of the sets in FSDKaggle2019. $^*$A few classes have slightly less than 75 clips.}
\vspace{+2mm}
\centering
\begin{tabular}{l|c|c|c}
\textbf{Aspect} 	& \textbf{curated train}   & \textbf{noisy train}  & \textbf{test}   \\
\hline
Clips/class	            & $\sim$75$^*$	    & 300        	   & $\sim$ 50 - 150 	\\
Total clips  	        & 4970 		        & 19,815 	       & 4481             \\
Labels/clip 	        & 1.2	            & 1.2 	           & 1.4     		\\
Clip length  	        & $\sim$0.3 - 30s  & $\sim$15s  	   &$\sim$0.3 - 30s              \\
Total duration  	    & $\sim$10.5h     & $\sim$80h  	       &$\sim$12.9h          \\
Labelling 	            & correct  	    & noisy 	        & exhaustive 		\\
       	                & (inexhaustive) 	 & 	        & 		\\

\end{tabular}
\label{tab:dataset_stats}
\vspace{-5mm}
\end{table}

\subsection{Noisy train set}
\label{ssec:noisy_train}

The noisy train set was prepared using the YFCC100M dataset \cite{YFCC100M}, which has the advantages of \textit{i)} being a very large and diverse dataset that is not correlated with Freesound in acoustics or domain, and \textit{ii)} offering permissive Creative Commons licenses that allow ease of use, modification, and redistribution. The original dataset contained $\sim$99M photos and $\sim$793k videos from $\sim$581K Flickr users. We dropped videos with licenses that disallowed making derivatives or commercial use, videos that were no longer available, and videos with audio decode errors that we could not transcode, leaving us with $\sim$201K 44.1 kHz mono WAV files. Video length varied with a maximum of 20 minutes, and a mean of $\sim$37s and median of $\sim$20s.

The Flickr video metadata (title, description, tags) proved to be too sparse to meaningfully map to our class vocabulary. Therefore, we used a content-based approach where we generated video-level predictions from a variety of pre-trained audio models: a shallow fully-connected network as well as variants of VGG and ResNet \cite{Hershey2017}, all of which were trained on a large collection of YouTube videos using the AudioSet class vocabulary. We generated sliding windows of $\sim$1s containing log mel spectrogram patches and aggregated the per-window predictions (using either maximum or average pooling) to produce a video-level vector of class scores. For each of our 80 classes, we kept the top 300 videos by predicted score for that class. We browsed the video labels and selected the maximum-pooled VGG-like model as producing a balance between reasonable predictions and a substantial amount of label noise. As a further source of noise, each final clip was produced by taking a random slice of a video of length up to 15 seconds (videos shorter than 15 seconds would be taken in their entirety).
Hence, the label noise can vary widely in amount and type depending on the class, including in- and out-of-vocabulary noises \cite{Fonseca2019learning}.

As listed in Table~\ref{tab:dataset_stats}, the noisy train set consists of 19,815 clips with a total of 24,000 labels (300 * 80). Labels per clip ranges from 1 to 7 with a mean of 1.2.
Clip length ranges from 1s to 15s (by construction), with a mean of 14.5s.
Therefore, the per-class training data distribution in FSDKaggle2019 is, for most of the classes, 300 clips from the noisy set and 75 clips from the curated set.
This means 80\% noisy / 20\% curated at the clip level, while at the duration level the proportion is more extreme considering the variable-length clips.
Since most of the train data come from YFCC, acoustic domain mismatch between the train and test set can be expected.
We conjecture this mismatch comes from a variety of reasons.
For example, through acoustic inspection of a small sample of both data sources, we find a higher percentage of high quality recordings in Freesound.
In addition, audio clips in Freesound are typically recorded with the purpose of capturing audio, which is not necessarily the case in YFCC.
\section{Baseline System}
\label{sec:baseline}
\subsection{Model Architecture and Training}

The baseline system uses a convolutional neural network that takes log mel spectrogram patches as input and produces predicted scores for 80 classes. We use an efficient MobileNet v1 \cite{howard2017mobilenets} architecture which lets us fit comfortably within the inference time limits of the challenge. Incoming audio (always 44.1 kHz mono) is divided into overlapping windows of size 1s with a hop of 0.5s. These windows are decomposed with a short-time Fourier transform using 25ms windows every 10ms. The resulting spectrogram is mapped into 96 mel-spaced frequency bins covering 20 Hz to 20 kHz, and the magnitude of each bin is log-transformed after adding a small offset to avoid numerical issues. The model consists of 1 convolutional layer followed by 13 separable convolution layers (which give MobileNets their compactness) followed by either max or average pooling, and an 80-way logistic classifier layer. The model contains $\sim$3.3M weights and, by comparison, is $\sim$8x smaller than a ResNet-50 while using $\sim$4x less compute. Detailed documentation is available in the public release of the baseline system code.\footnote{\url{https://github.com/DCASE-REPO/dcase2019_task2_baseline}}

To combat label noise, we use dropout (inserted before the classifier layer) as well as label smoothing \cite{inceptionv3} which replaces each label's target value with a hyperparameter-controlled blend of the original value and 0.5 (representing a uniform probability distribution). To combat the domain mismatch between the test set and noisy train set, we use transfer learning by training a model on the noisy set first to learn a representation, and then use a checkpoint from that run to warm-start training on the curated train set. In addition, we used batch normalization, exponential learning rate decay, and the Adam optimizer. We trained models on the curated data alone, the noisy data alone, the curated and noisy combined, noisy first followed by warm-started curated, as well as a weighted version of warm-started training that we describe next. 

\subsection{Results}

Table \ref{baseline_results_table} shows the {\lwlrap} values produced by our various baseline models when evaluated on the entire test set (i.e., including both the public and private splits). Each row lists the best {\lwlrap} obtained from a small grid search (using the public test set for evaluation) that varied maximum vs average pooling, learning rate, label smoothing, dropout, learning rate decay, and whether or not we used batch normalization. The baseline system that we settled on was ``Warm-started curated", which achieved a {\lwlrap} of 0.537 on the public test set (see the publicly released baseline code for hyperparameter choices).

\vspace{-4mm}
\begin{table}[h!]
\caption{Baseline system results on the \textit{entire} test set.}
\vspace{+2mm}
\centering
\begin{tabular}{l|l}
\textbf{Training approach} & \textbf{\lwlrap} \\ \hline
Curated only      &     0.542       \\ 
Noisy only        &     0.312       \\ 
Curated + Noisy   &     0.522       \\ 
\textbf{Warm-started curated} &  \textbf{0.546}       \\ 
Weighted warm-started curated & 0.561  \\ 
\end{tabular}
\label{baseline_results_table}
\end{table}

Comparing ``Noisy only" and ``Curated + Noisy" to ``Curated only" shows a considerable domain mismatch where we hurt our performance when we blindly add more data. A transfer learning approach of warm-starting the curated training with a noisily trained model gives us a small boost in performance. 

We conducted a class-based analysis of the results in a public Colab notebook\footnote{\url{https://colab.research.google.com/drive/1vIpAWCGSHyYYLblbtMzAKLNTkHcr1SMK}} where we look at the best and worst classes of each model.
The baseline system attains highest {\lwlrap} values for \emph{Bicycle bell} (0.894) and \emph{Purr} (0.873); lowest values occur for \emph{Cupboard open or close} (0.219) and \emph{Chirp, tweet} (0.127).

We also analyse the correlations between various pairs of models. It becomes evident that, at least for our baseline models, there are classes where using curated data alone is better while there are other classes where the noisy data is better. One simple way to incorporate this in training is to use the ratio of noisy to curated {\lwlrap s} as a per-class weight during noisy training to boost classes that have value and suppress classes that do not. When we warmstart with this weighted noisy model, we get a further boost in performance. This optimization is not included in the released baseline. 

\section{Conclusion}
\label{sec:conclusion}

In this paper, we have described the task setup, dataset, and baseline system of DCASE2019 Task 2 ``Audio tagging with noisy labels and minimal supervision''. 
This task was hosted on the Kaggle platform as ``Freesound Audio Tagging 2019''.
The goal is to adequately exploit a large set of noisy-labeled data and a small quantity of manually-labeled data, in a multi-label audio tagging setting with a vocabulary of 80 everyday sound classes.
In addition, the dataset poses an acoustic mismatch problem between the noisy train set and the test set due to the fact that they come from different web audio sources.
We believe this can correspond to a realistic scenario given the difficulty in gathering large amounts of manually labeled data.
{\lwlrap} is proposed as evaluation metric. Baseline experiments indicate that leveraging noisy-labeled data with a distribution shift for sound event tagging can be challenging.
The FSDKaggle2019 dataset and the baseline system proposed are freely available and not limited for use within the competition.


\section{ACKNOWLEDGMENT}
\label{sec:ack}
We thank Addison Howard and Walter Reade of Kaggle for their invaluable assistance with the task design and Kaggle platform, and everyone who contributed to FSDKaggle2019 with annotations, especially Jordi Pons and Xavier Favory.
Eduardo Fonseca is also grateful for the GPU donated by NVidia.

\newpage
\bibliographystyle{IEEEtran}
\bibliography{refs}
%
%
%
%
%
%
%
%
%

\end{sloppy}
\end{document}